# The reference frames of Mercury after MESSENGER

Alexander Stark [1], Jürgen Oberst [1,2], Frank Preusker [1], Steffi Burmeister [2], Gregor Steinbrügge [1], Hauke Hussmann [1]

[1] German Aerospace Center, Institute of Planetary Research, Berlin, Germany

[2] Technische Universität Berlin, Institute of Geodesy and Geoinformation Sciences, Berlin, Germany

**Abstract**

We report on recent refinements and the current status for the rotational state models and the reference frame of the planet Mercury. We summarize the performed measurements of Mercury rotation based on terrestrial radar observations as well as data from the Mariner 10 and the MESSENGER missions. Further, we describe the different available definitions of reference systems for Mercury, which are realized using data obtained by instruments on board MESSENGER. In particular, we discuss the dynamical frame, the principal-axes frame, the ellipsoid frame, as well as the cartographic frame. We also describe the reference frame adopted by the MESSENGER science team for the release of their cartographic products and we provide expressions for transformations from this frame to the other reference frames.

## 1. Introduction

The availability of a planetary reference frame and coordinate knowledge is critical for any remote sensing application of a celestial body. Thereby, a reference frame is a realization of a defined reference system. Of particular importance is the body-fixed reference system as it provides time-invariant coordinates for surface features. The body-fixed reference system is defined by a transformation operation from an inertial reference frame with the help of a rotation model and the definition of a basic datum (i.e. a prime meridian). The resulting body-fixed frame, or cartographic frame, is obtained from measurement of the rotational state and the selection of a surface feature to be located at the prime meridian.

The reference frame of Earth is known to very high accuracy (see other papers of the special issue). Benefiting from decades of lunar laser ranging the reference frame of Earth's Moon is also known to an accuracy of some meters (see other papers of the special issue). The reference frame of Mars was studied by several spacecraft and landers, but still demands further improvement in view of possible future human exploration. Venus and Mercury, though, exhibit relatively poor knowledge of their reference frames. While for Venus the definition of a reference frame suffers from lacking of detailed surface observations, in the case of Mercury the absence of orbital observations was a major shortcoming.

The poor knowledge on the planet does not state that the planet was not in the focus of research in planetary science. With the development of optical instruments astronomers made efforts to identify and track surface features in order to measure the rotation parameters of Mercury. Owing to the planet's proximity to the Sun, it was believed that Mercury's rotation state is synchronized with the planet's orbital motion (i.e., the rotation period to be equal to the orbital period), as was known to be in the case of the Moon. However, radar observations by Pettengill and Dyce (1965) revealed that Mercury's rotation period is 2/3 of its orbital period, obviously a resonance between spin and orbital motion (Colombo, 1965). The presence of Mercury's spin-orbit resonance allows the definition of a dynamical reference system, which is well constrained by orbital dynamics of the planet.

Data by the Mercury Surface, Space ENvironment, GEochemisty, and Ranging (MESSENGER) spacecraft (Solomon et al., 2011) significantly improved the knowledge on the innermost planet of our Solar System. Accordingly, the reference frame of Mercury, currently based on Earth-based radar observations should be revised.

In this paper, we first summarize existing rotation models of Mercury including the resonant rotation model and rotation models obtained from measurements. Next we discuss possible definitions of Mercury's reference systems and provide the corresponding frames. In particular, we present the reference frame adopted by the MESSENGER science team for the release of their cartographic products. Furthermore, we provide expressions for transformations between frames. The paper concludes with a discussion and suggestions for further improvements of Mercury's reference frame.

## 2. Mercury rotation models

A rotation model of a celestial body is characterized by three time-dependent angles: declination $\delta$, right ascension $\alpha$ and prime meridian angle $W$ (Archinal et al., 2011). While the first two specify the orientation of the rotation axis and its precession and nutation, the latter describes the rotation about the rotation axis including the libration in longitude. Given these three rotation angles a transformation between inertial and body-fixed coordinates for any given time can be constructed. All three angles are typically expressed in the form of analytic expressions decomposed in a secular component and a summation of trigonometric functions.

Thereby the secular component is given by an initial orientation at the J2000.0 epoch and a power series in time. As the rotational behavior typically extends the available observation time some rotation parameters are computed based on the orbital motion of the body (e.g. long-period precession rates of the rotation axis).

For Mercury with its 3:2 spin-orbit resonance the rotation model can be defined based on the orbital motion of the planet (see section 2.1). These computed values can be used for comparisons of actual measurements of Mercury's rotation (see section 2.2).

*2.1. Resonant rotation model*

The resonant rotation model of Mercury is based on the two key assumptions - that the rotation rate is firmly tied to the planet's mean orbital motion (including pericenter precession) and that the rotation axis occupies a Cassini state. The former is a consequence of the 3:2 spin-orbit resonance, implying that the planet rotates three times, as it orbits the Sun twice. The latter assumption implies that the rotation axis lies always in the plane spanned by the normal vectors of the orbit plane and the Laplace plane. The normal of the Laplace plane (or invariable plane) is the axis about which the orbital plane precesses with a constant inclination.

The parameters of the resonant rotation model may be obtained from observations of Mercury's orbital motion, as can be conveniently extracted from its osculating orbital elements in published Solar System ephemeris data, e.g. Jet Propulsion Laboratory Development Ephemeris (Folkner et al., 2014). When averaged over sufficiently long time intervals, the periodic parts in the time series of the orbital elements vanish, with the mean orbital elements remaining. Thus the averaging of the orbital elements is performed by a decomposition of the time series in a secular and a periodic part. For the resonant rotation model we use the decomposition of Mercury's orbital elements reported by A. Stark, Oberst, and Hussmann (2015). Since the obliquity is very small (about 2 arc minutes) and the precession periods are very long (hundreds of thousands of years) we perform a linearization of the model in first order in obliquity $\varepsilon_\Omega$ and first order in time *t*. The obliquity is connected to the interior structure of Mercury, in particular to the normalized polar moment of inertia *C/MR²*.

Using the secular parts of the osculating orbital elements obtained by A. Stark, Oberst, and Hussmann (2015) the Cassini state declination $\delta^{CS}$, right ascension $\alpha^{CS}$ and prime meridian angle $W^{CS}$ (with respect to the International Celestial Reference Frame (ICRF)) is given by

$$\delta^{CS}(t) = 61.44780272° - 0.95540886°\varepsilon_\Omega/° + (-0.00484640° - 0.00041197°\varepsilon_\Omega/°)t/\text{cy} \tag{1}$$

$$\alpha^{CS}(t) = 280.98797069° + 0.61780624°\varepsilon_\Omega/° + (-0.03280760° - 0.00288486°\varepsilon_\Omega/°)t/\text{cy} \tag{2}$$

$$W^{CS}(t) = 329.75640656° - 0.54266991°\varepsilon_\Omega/° - W_{\text{lib}}(0) + \\ (6.138506839° + 7.01 \cdot 10^{-8}°\varepsilon_\Omega/°)t/\text{d} + W_{\text{lib}}(t). \tag{3}$$

Thereby, the $t$ is the time from the J2000.0 epoch, which is measured in Julian centuries (cy) in case of $\delta^{CS}$ and $\alpha^{CS}$ and in days (d) for $W^{CS}$. The obliquity $\varepsilon_\Omega$ is measured in degrees and the term $W_{\text{lib}}(t)$ denotes the longitudinal libration terms. It should be stressed that the obliquity modifies the precession rates as well as the prime meridian constant (see section 3.4).

With the help of equations 1 to 3 and measurements for the obliquity and the libration amplitude the rotational state of Mercury is fully constrained. Recently, Baland et al. (2017) have extended the Cassini state model to account for pericenter precession and tidal deformation of Mercury. We discuss the incorporation of this model in Appendix A. With expressions obtained there the resonant rotation model can be brought to agreement to measured orientations of the rotation axis, as in this model the strict requirement of co-planarity is relaxed.

## 2.2. Measured rotation parameters

Table 1 provides an overview on the measured rotational parameters. Thereby the orientation of the rotation axis is parameterized by the declination $\delta(t) = \delta_0 + \delta_1 t/\text{cy}$ and right ascension $\alpha(t) = \alpha_0 + \alpha_1 t/\text{cy}$. The rotation about that axis is defined by the prime meridian angle $W(t) = W_0 + W_1 t/\text{d} + W_{\text{lib}}(t)$ and $A_{\text{lib}} = \max_t W_{\text{lib}}(t)$ is the amplitude of the forced libration on longitude $W_{\text{lib}}(t)$. [The performed measurements are discussed chronologically in the text below.]

The first measurements of Mercury's rotation were carried out by visual telescopic observations, which suggested that Mercury was in a 1:1 spin-orbit resonance (Lowell, 1902; Schiaparelli, 1890). In contrast, radar observations by Pettengill and Dyce (1965) provided first evidence that Mercury is trapped in a 3:2 spin-orbit resonance. A detailed analysis of their measurements was provided by Dyce et al. (1967). Following to that pioneering observation McGovern et al. (1965) and Smith and Reese (1968) demonstrated that most of the early telescopic observations were also in agreement with a 3:2 spin-orbit resonance. Incorporating new observations Camichel and Dollfus (1968) confirmed this conclusion. Pettengill and Dyce (1965) could not obtain an estimate for the orientation of the rotation axis but claimed that the axis is approximately normal to the orbital plane of Mercury.

In 1970, the commission for Physical Study of Planets and Satellites of the International Astronomical Union (IAU) adopted and recommended the use of a "provisional" first rotation model (Hall et al., 1971), which included a rotation period of 58.6462 days and a rotation axis normal to Mercury's orbital plane of the 1950.0 epoch. Thereby the rotation period is based on computation by Colombo (1965), who assumed a perfect 3:2 resonance of Mercury's rotation to its orbit. It is worth noting that the adopted rotation model (when transformed to the ICRF), was already in good agreement with recent computation for Mercury's orbit normal and the resonant rotation rate by A. Stark, Oberst, and Hussmann (2015). However, presumably due to rounding effects, the initial precision was lost through the years. Indeed, the first report of the IAU Working Group for Cartographic Coordinates and Rotational Elements (WGCCRE) (Davies et al., 1980) gives the rotation axis coordinates with a precision of only 6 arc min (0.1 degree). The provisionally adopted rotational rate was in use for nearly 50 years and was revised only recently, following computations of A. Stark, Oberst, and Hussmann (2015) and measurements by Mazarico et al. (2014) and A. Stark, Oberst, Preusker, et al. (2015).

The first spacecraft observations of Mercury were performed by Mariner 10 during its three flybys in 1974 and 1975. The data collected by the spacecraft provided image coverage for about half of the planet's surface. However, the Mariner 10 data contributed only marginally to the knowledge of Mercury's rotation. Through an analysis of the images obtained by Mariner 10 Klaasen (1975, 1976) reported on the first space-based measurements of the rotation rate and the orientation of the axis. However, due to poor knowledge of camera characteristics (focal length and orientation parameters) the accuracy of these estimates was rather limited and could not improve over the resonant rotation model of that time.

The first accurate measurements of Mercury rotation parameters, including rotation axis orientation and librations were made by Margot et al. (2007) using Earth-based radar observations, which were updated later using the same technique, but longer radar observation sequences of 10 years (Margot et al., 2012). Also, Margot (2009) revised the orientation and the precession rates of the orbit plane normal.

In 2011 the MESSENGER mission entered orbit about Mercury and provided new information on the rotation of the planet from two different measurement techniques. Parameters of Mercury's rotational state were revised using radio science (Mazarico et al., 2014; Verma & Margot, 2016) and co-registration of laser altimeter tracks with respect to image data (A. Stark, Oberst, Preusker, et al., 2015). Mazarico et al. (2014) provided estimates for the orientation of the rotation axis and the rotation period. Verma and Margot (2016) also used radio science data but obtained measurements of the rotation axis which disagree by 13 arc seconds with those of Mazarico et al. (2014). A. Stark, Oberst, Preusker, et al. (2015) obtained estimates for the orientation of the rotation axis, the rotation rate and the libration amplitude. Their observations are referenced to MJD56353.5 (about midterm of the MESSENGER mission) and they have used the rotation axis precession rates reported by A. Stark, Oberst, Preusker, et al. (2015) to obtain the orientation of the rotation axis at the J2000.0 epoch.

Hence, at present time there are four recent measurements of rotation axis orientation (Margot et al., 2012; Mazarico et al., 2014; A. Stark, Oberst, Preusker, et al., 2015; Verma & Margot, 2016), two measurements of the rotation rate (Mazarico et al., 2014; A. Stark, Oberst, Preusker, et al., 2015) and two measurements of the libration amplitude (Margot et al., 2012; A. Stark, Oberst, Preusker, et al., 2015). Thereby the estimates by (Margot et al., 2012), A. Stark, Oberst, Preusker, et al. (2015) and Verma and Margot (2016) agree within their respective uncertainties. Likewise, the libration amplitude measurements by Margot et al. (2012) and A. Stark, Oberst, Preusker, et al. (2015) are in good agreement within their reported errors [and differ by only 0.3 arc seconds]. In contrast, the rotation rate measurements of Mazarico et al. (2014) and A. Stark, Oberst, Preusker, et al. (2015) differ significantly, by about 9 meters at the equator (or 6 seconds) after one Mercury rotation.

## 3. Mercury reference frames

### 3.1. Pre-MESSENGER reference frames

Shortly after the observation of the 3:2 spin-resonance of Mercury the first definition of Mercury's body-fixed reference system was performed. In 1970 the commission for Physical Study of Planets and Satellites of the IAU realized the first reference frame by adopting the a rotation model of Mercury and the definition of the prime meridian by the subsolar point at the first perihelion passage of 1950 (J.D. 2433292.63) (Hall et al., 1971). Only some years later the Mariner 10 mission provided first images of Mercury's surface and a feature-based definition of the prime meridian became possible (Davies & Batson, 1975). Since the prime meridian of the 1970 definition was on the night side of Mercury during the Mariner 10 flybys Murray et al. (1974) defined the longitude 20° W by the small crater Hun Kal which was named after the Mayan numeral for 20 (cf. Fig.1). Following this definition the IAU WGCCRE published a refined reference frame of Mercury in their first report (Davies et al., 1980). While the rotation axis and the rotation rate were devised from the orbital dynamics, the prime meridian constant $W_0$ was now determined by the crater Hun Kal. In the following years with the help of the improved control point solution based on Mariner 10 images the prime meridian constant was revised (Davies et al., 1996; Davies et al., 1983; Robinson et al., 1999). As the data obtained by MESSENGER significantly improved the knowledge on the rotation state of Mercury, we propose a revision of the reference frame parameters adopted by the IAU WGCCRE.

*3.2. MESSENGER reference frame*

The MESSENGER science team adopted a new common reference frame to support the handling of all the mission's data products in 2015 (Perry & McNut, 2015). The rotation parameters were combined from Earth-based radar and MESSENGER radio science measurements as well as from a fit to the ephemeris of Mercury. As the rotation rate changed significantly compared to the previous model, a corresponding adjustment of the prime meridian constant became mandatory. This was accomplished by observations of the location of the crater Hun Kal with respect to the inertial frame. The Mercury Dual Imaging System (MDIS) acquired about 12 images with resolutions less than 1 km in which the crater Hun Kal was identifiable. Errors in spacecraft orbit, camera calibration and pointing, as well as effects of different resolution and visibility conditions were met by averaging and weighting the observations (A Stark, 2015). In the analysis by A Stark (2015), Hun Kal was found to be offset by approximately 0.09° (3.9 km) from -20°E when the rotation model of Archinal et al. (2011) was used. With the rotation rate measured by Mazarico et al. (2014) and the libration amplitude and obliquity measurements of Margot et al. (2012) the prime meridian constant was estimated to $W_0^{\text{MSGR}} = 329.5988 \pm 0.0037°$. This value differs by 0.0519° (~2.2 km) from the value adopted for Mercury by the IAU in their report of 2009 (Archinal et al., 2011). The reference frame including rotational parameters adopted for MESSENGER cartographic products is highlighted in bold face in Table 1.

*3.3. Cartographic frame*

Preusker et al. (2017) computed a stereo digital terrain model (DTM) of the H-6 (Kuiper) quadrangle of Mercury (288 to 360° E and 22.5°S to 22.5°N). The authors used approximately 10,500 MDIS images and performed a photogrammetric block adjustment, which improved the pointing knowledge of the camera. As a result a geometrically stable terrain model with a resolution of 222 meters per pixel was obtained. A comparison to profiles collected by the Mercury Laser Altimeter (MLA) showed very good agreement between the two data sets.

Although, the small crater Hun Kal is not identifiable in the DTM it can be observed in five MDIS images used for computation of the DTM and corrected in the process of the block adjustment. The authors report 0.465° S, 339.995° E as coordinates for Hun Kal in the MESSENGER reference frame (see section 3.2). The derived offset to 340° E (20° W) is only 0.0052° or 220 m (about one DTM pixel). In order to restore the longitude of Hun Kal to 20° W one has to revise the prime meridian constant to $W_0^{\text{H6 DTM}} = 329.6040 \pm 0.0052°$ (where the error is taken as the size of one DTM pixel).

*3.4. Dynamical frame*

As Mercury is in a 3:2 spin-orbit resonance, it is also possible to define a reference system where the prime meridian is oriented towards the Sun every second passage through the pericenter. In the resonant rotation model derived in section 2.1, the prime meridian constant is given such that Mercury points opposite to the Sun direction at the last pericenter passage just prior the J2000.0 epoch. In particular, this pericenter passage occurs 42.71274 days before the J2000.0 epoch (A. Stark, Oberst, & Hussmann, 2015). We want to stress that this definition of the prime meridian is based on averaged (secular) orbital elements and thus the actual pericenter passage may differ by about 1 minute from the specified epoch due to variations in Mercury's orbit.

An effect neglected previously is the displacement of the prime meridian constant by the obliquity. Due to Mercury's obliquity the prime meridian constant has to be modified in order to comply with the definition that the prime meridian is oriented to the Sun every second pericenter passage. In fact, the location of the prime meridian on the surface of Mercury changes by 6.4 meters with a change of the obliquity $\varepsilon_\Omega$ by 1 arc second. Given the measured value for the obliquity of 2.029 ± 0.085 arc minutes and the libration amplitude of 38.9 ± 1.3 arc seconds (A. Stark, Oberst, Preusker, et al., 2015) we obtain $W_0^{CS} = 329.7369° \pm 0.0052°$ (the error bar is obtained through error propagation of uncertainty of the averaged orbital elements (A. Stark, Oberst, & Hussmann, 2015) and the uncertainty of the obliquity and libration amplitude). The derived value is 0.0184° (780 meters at the equator) away from prime meridian defined without taking into account the obliquity effect on the prime meridian constant.

In order to transform from the reference frame adopted for MESSENGER data to the dynamical frame (with the obliquity and libration values from Margot et al. (2012)) one has to rotate the x-axis counter-clockwise about the rotation axis by $0.1380° - 0.1446°\ t/\text{cy}$. At the midterm of the MESSENGER mission (MJD56353.5) this angle amounts to 0.12° or 5 km. Within the transformation derived here we neglected the small deviation (about 1 arc second) of the rotation axis from the complanarity condition of the nominal Cassini state. A dynamical frame based on the extended Cassini state is provided in the Appendix A.

*3.5. Principal-axes frame*

The principal-axes reference system is defined by the orientation of the principal components of Mercury's moments of inertia. The origin of the principal-axes reference system is the center of mass. As the gravity field reflects the mass distribution it can be used to derive the orientation of the principal axes and thus to obtain a principal-axes frame. Thus, we consider the degree-2 coefficients of an expansion of the gravity field in spherical harmonics and use the most recent estimates provided by Verma and Margot (2016). As the authors solved for the orientation of the rotation axis in their inversion we transformed their measurements to the MESSENGER reference frame (see section 3.2). Thereby we did not consider any effects due to the adopted rotation rate of Mercury and used the reference epoch of J2000.0. The obtained transformation matrix is given by

$$P = R_x(52 \pm 87") \cdot R_y(42 \pm 45") \cdot R_z(70 \pm 108"),$$

where the error bars of the angles were computed based on the adopted error bars for the gravity field coefficients in Verma and Margot (2016). $R_{x,y,z}$ are rotation matrices describing a rotation using the right-hand rule about the respective axis.

Based on the obtained values we conclude that, as expected, the rotation axis coincides with the axis of largest moment of inertia within the limited accuracy of the latter. The axis of smallest inertia, however, should be aligned with the Mercury-Sun direction at the pericenter passage. Thus, the principal-axis frame should coincide with the dynamical frame and one would expect a deviation in the order of 430 arc seconds, since it was shown that the MESSENGER reference frame is offset by 0.12° from the dynamical frame. The observed offset of 70 arc seconds, however, is somewhat smaller and inconsistent with the dynamical frame. One possible reason for this discrepancy is that the offset is strongly correlated with the rotation rate. Verma and Margot (2016) used the outdated rotation rate of 6.1385025°/day, while the usage of the resonant rotation rate (see section 2.1) could lead to more consistent estimates. Furthermore, the poor knowledge of the gravity field in the southern hemisphere of Mercury could lead to biased estimates of the degree-2 gravity field coefficients.

*3.6. Ellipsoid frame*

The shape of Mercury can be approximated by a tri-axial ellipsoid which provides a definition of an ellipsoid reference system. Thereby the origin of the ellipsoid reference system (i.e. the center of figure) can have a deviation from the origin of the dynamical or principal-axis reference systems (i.e. the center of mass). This offset between the two reference systems may hint at asymmetries in mass distribution within the planet.

While early estimates on the shape of Mercury were based on Earth-based radar observations (Anderson et al., 1996) the data provided by MESSENGER significantly improved the knowledge on the global shape of the planet. Indeed, ellipsoid parameters were obtained using MDIS limb images (Elgner et al., 2014), as well as from MLA profiles combined with MESSENGER radio link occultation data (Perry et al., 2015). In addition, we derive the shape ellipsoid from a MDIS DTM based on stereo images (Becker et al., 2016) by computing the degree-2 coefficients of an expansion of the topography in spherical harmonics (see summary in Table 2). While the ellipsoid parameters obtained by Perry et al. (2015) and the ones based on the MDIS DTM are more or less consistent, the ellipsoid characteristics obtained from limb images deviate remarkably, probably due to MDIS calibration issues early in the mission. Thus, for the definition of the ellipsoidal frame we average the values from Perry et al. (2015) and from the MDIS DTM and set the error bars to contain both estimates (last column of Table 2).

The transformation from the MESSENGER frame to the ellipsoid frame is then given by a translation of

$$(dx, dy, dz) = (0.059 \pm 0.017, 0.127 \pm 0.007, -0.0675 \pm 0.030) \text{ km}$$

and a subsequent rotation by

$$E = R_x(0.91 \pm 1.79°) \cdot R_y(-2.67 \pm 0.75°) \cdot R_z(15.8 \pm 0.7°).$$

While the orientation of the short axis is consistent with the rotation axis orientation, the offset of the long axis with respect to the dynamical (section 3.4) and principal axes frame (section 3.5) is puzzling and requires further investigations.

## 4. Discussion

The large deviation between the dynamical frame and the feature-based MESSENGER frame is an important issue. Due to the spin-orbit resonance it can be assumed that the dynamical reference system should coincide with the reference system defined by the principal moment of inertia of Mercury (see section 3.5). Thus, the observed deviation of the dynamical frame would also hold for the frame defined by the principal axes. When the MESSENGER frame is used certain coefficients ($C_{21}$ and $S_{21}$) of the expansion of Mercury's gravity in terms of spherical harmonics could not be assumed to be vanishing. As a consequence, the inversion of Mercury's gravity field could be compromised.

The MESSENGER mission ended in April 2015. However, the analysis of MESSENGER data is still ongoing and new refinements of Mercury's reference frame are still expected. The observed longitudinal offset of the ellipsoidal and principal-axis frame requires further investigations. Further improvements in Mercury's rotation models and reference frame will come with ESA's Bepi Colombo mission, which is readied for launch in 2018. The spacecraft, equipped with a camera system and a powerful laser altimeter, will provide uniform global coverage of the shape and the gravity field. The rotational dynamics will also be determined with a high level of accuracy (Imperi et al., 2017).

With the improved knowledge on Mercury's rotational state a re-definition of the prime meridian will become mandatory. Given the diameter of the crater Hun Kal of about 1.5 km the selection of a smaller feature will allow a more precise definition of the cartographic reference frame. The definition of the prime meridian based on orbital dynamics of Mercury has the advantage that it is independent on the adopted rotation rate for Mercury. However, such a definition could hardly be checked as no observations of Mercury are available at the reference epoch J2000.0. One possible solution would be to use a different reference epoch (within the MESSENGER observation period) and consistently tie the dynamical prime meridian to a surface feature. Besides the rotation rate independent definition, this would also have the advantage that the dynamical and cartographic frames would be consistent within their level of uncertainty. The disadvantage is that Hun Kal would be offset by about 5 km from the 20° W longitude.

**Acknowledgements**

This research was funded by a grant from the German Research Foundation (OB124/11-1). A. Stark was funded by a research grant from the Helmholtz Association and German Aerospace Center (DLR). J. Oberst gratefully acknowledges being hosted by the Moscow State University of Geodesy and Cartography (MIIGAiK). The authors thank all members of the MESSENGER science and instrument teams.

**Appendix A – Extended dynamical frame**

Recently, Baland et al. (2017) have extended the Cassini state model to account for pericenter precession and tidal deformation of Mercury. Thereby the authors express the orientation of Mercury with respect to its Laplace plane. For the transformation to the ICRF from the reference frame defined by the Laplace plane normal and the node of Laplace plane and ICRF equator (also used by Baland et al. (2017)) we use the following transformation matrix

$$\begin{pmatrix} 0.06140463088547411 & 0.9978489634266365 & 0.02295468349172782 \\ -0.9346751836048649 & 0.06555490971991894 & -0.3494064323460937 \\ -0.35015963853509846 & 0 & 0.9366900381347978 \end{pmatrix},$$

which is based on the orbital elements of A. Stark, Oberst, and Hussmann (2015). In particular, the ICRF spherical coordinates of the Laplace pole are given by $(69.5029204°, 273.7587151°)$.

We express the rotational angles as functions of the precession amplitude $\varepsilon_\Omega^{k_2}$, nutation amplitude $\varepsilon_\omega^{k_2}$ and the tidal deviation amplitude $\varepsilon_\zeta$ (see Eq. 64 to 66 of Baland et al. (2017)). These amplitudes are connected to the interior structure of Mercury, in particular to the normalized polar moment of inertia $C/MR^2$, the tidal Love number $k_2$ and the tidal quality factor $Q$. Hence, the Cassini state declination $\delta^{\text{eCS}}$, right ascension $\alpha^{\text{eCS}}$ and prime meridian angle $W^{\text{eCS}}$ with respect to the ICRF are

$$\delta^{\text{eCS}}(t) = 61.44780272° - 0.95540886°\varepsilon_\Omega^{k_2}/° + 0.46675751°\varepsilon_\omega^{k_2}/° + 0.2952861°\varepsilon_\zeta/°$$
$$+ (-0.00484640° - 0.00041197°\varepsilon_\Omega^{k_2}/° + 0.00694873°\varepsilon_\omega^{k_2}/°$$
$$- 0.00133294°\varepsilon_\zeta/°)t/\text{cy} + 0.00001960°(t/\text{cy})^2$$

(4)

$$\alpha^{\text{eCS}}(t) = 280.98797069° + 0.61780624°\varepsilon_\Omega^{k_2}/° + 1.84941502°\varepsilon_\omega^{k_2}/° + 1.99893401°\varepsilon_\zeta/°$$
$$+ (-0.03280760° - 0.00288486°\varepsilon_\Omega^{k_2}/° - 0.00805508°\varepsilon_\omega^{k_2}/°$$
$$+ 0.00055120°\varepsilon_\zeta/°)t/\text{cy} - 0.00002449°(t/\text{cy})^2$$

(5)

$$W^{\text{eCS}}(t) = 329.75640656° - 0.54266991°\varepsilon_\Omega^{k_2}/° - W_{\text{lib}}(0) - 1.62449296°\varepsilon_\omega^{k_2}/° -$$
$$1.7558277°\varepsilon_\zeta/° + (6.138506839° + 7.01 \times 10^{-8}°\varepsilon_\Omega^{k_2}/° + 19.58 \times 10^{-8}°\varepsilon_\omega^{k_2}/° -$$
$$1.10 \times 10^{-8}°\varepsilon_\zeta/°)t/\text{d} + W_{\text{lib}}(t).$$

(6)

Thereby, $t$ is the time and is measured in centuries (cy) (in case of $\delta^{\text{eCS}}$ and $\alpha^{\text{eCS}}$) and in days (d) (for $W^{\text{eCS}}$). The three $\varepsilon_\Omega^{k_2}$, $\varepsilon_\omega^{k_2}$ and $\varepsilon_\zeta$ obliquity parameters are measured in degrees and the term $W_{\text{lib}}(t)$ denotes the longitudinal libration terms. For the case of a rigid Mercury ($k_2 \to 0$) and neglecting the effect of the pericenter precession ($\varepsilon_\omega^{k_2} \to 0$ and $\varepsilon_\zeta \to 0$) the obliquity parameter $\varepsilon_\Omega^{k_2}$ becomes the Cassini state obliquity $\varepsilon_\Omega$ and coincides with equations 1 to 3 in section 2.1.

With the help of the provided equations and given an observation of Mercury's rotation axis orientation at a specific epoch $t'$ and an independent measurement of the tidal Love number $k_2$ it is possible to solve for the normalized polar moment of inertia $C/MR^2$ and the tidal quality factor $Q$. Furthermore, once the parameters $\varepsilon_\Omega^{k_2}$, $\varepsilon_\omega^{k_2}$ and $\varepsilon_\zeta$ are determined it is straightforward to derive the orientation and precession rate of the rotation axis at the J2000.0 epoch (Baland et al., 2017). Using the observations for the rotation axis orientation of A. Stark, Oberst, Preusker, et al. (2015) at MJD56353.5 and $k_2 = 0.5 \pm 0.1$ Baland et al. (2017) have obtained $C/MR^2 =$ $0.3433 \pm 0.0134$ and $Q = 89 \pm 261$. The corresponding amplitudes are $\varepsilon_\Omega^{k_2} = 2.032 \pm 0.080$ arc minutes, $\varepsilon_\omega^{k_2} = 0.868 \pm 0.034$ arc seconds and $\varepsilon_\zeta = 0.995 \pm 2.914$ arc seconds.

With the extended Cassini state model the realization of the extended dynamical reference system is possilbe. In this reference system the z-axis coincides exactly with rotation axis. Given the values for the precession and nutation amplitudes obtained by Baland et al. (2017) one obtains $W_0^{eCS} = 329.7360° \pm 0.0053°$ (the error bar is obtained through error propagation of uncertainty of the averaged orbital elements (A. Stark, Oberst, & Hussmann, 2015) and measurement uncertainty of the obliquity (Baland et al., 2017)).

**Figures and Tables**

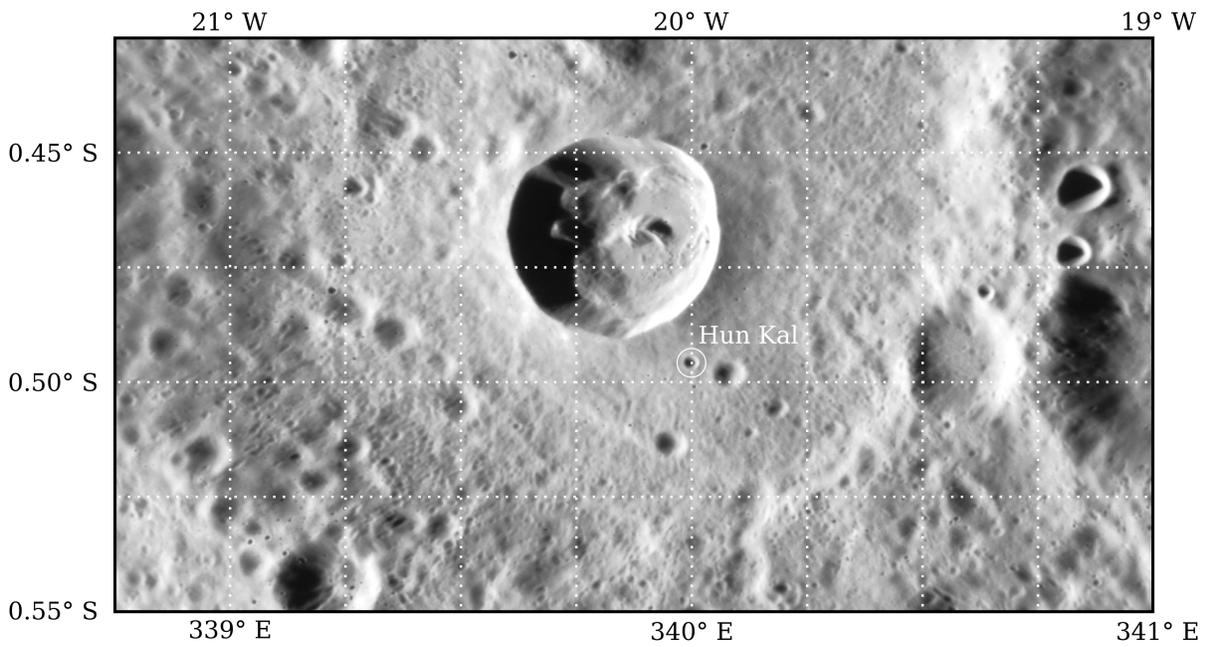

Figure 1: Part of a MDIS NAC image (EN1007130467M) showing Hun Kal (indicated by a circle). The ground resolution of the image is 73 m.

Table 1: Overview about observations of Mercury's rotation.

| Author | $\alpha_0$ [°] | $\alpha_1$ [°/cy] | $\delta_0$ [°] | $\delta_1$ [°/cy] | $W_0$ [°] | $W_1$ [°/day] | $A_{lib}$ ["] |
|---|---|---|---|---|---|---|---|
| Pettengill and Dyce (1965) (radar) | | | | | | 6.1±0.5 | |
| Colombo (1965) (computation) | | | | | | 6.138505138 | |
| McGovern et al. (1965) (visual) | | | | | | 6.16±0.04 | |
| Dyce et al. (1967) (radar) | | | | | | 6.1±0.3 | |
| Camichel and Dollfus (1968) (visual) | | | | | | 6.136 ± 0.003 | |
| Smith and Reese (1968) (visual) | | | | | | 6.1367 ± 0.0021 | |
| IAU 1970 (computation, Davies and Batson (1975)) | 280.980 | | 61.447 | | 329.714 | 6.1385025 | |
| Murray et al. (1972) (visual) | | | | | | 6.1387 ± 0.0009 | |
| Klaasen (1975) (Mariner 10) | | | | | | 6.1369 ± 0.0018 | |
| Klaasen (1976) (Mariner 10) | 281±10 | | 63±5 | | 329.68 | 6.13852 ± 0.00052 | |
| Davies et al. (1980); Davies et al. (1983) (computation, $W_0$ from Mariner 10 images) Davies et al. (1996) (Mariner 10) | 281.01 | −0.0033 | 61.45 | −0.005 | 329.71 | 6.1385025 | |
| Robinson et al. (1999) (Mariner 10) | | | | | 329.548 ± 0.470 | | |
| Margot et al. (2007) (Earth-based radar) | 281.0097 ± 0.0054 | | 61.4143 ± 0.0021 | | | | |
| Margot (2009) (computation) | 280.9880 | **−0.00328** | 61.4478 | **−0.0049** | 329.75 | | |
| Margot et al. (2012) (Earth based-radar) | **281.0103** | | **61.4155** | | | | |
| Mazarico et al. (2014) (MESSENGER radio science) | 281.00480 ± 0.00480 | | 61.41436 ± 0.0021 | | | | 35.8 ± 2 |
| Stark et al. (2015a) (computation) | 280.987971 ± 0.000099 | −0.032808 ± 0.000020 | 61.447803 ± 0.000036 | −0.0048464 ± 0.0000073 | 329.7564 ± 0.0051 | 6.138506839 ± 0.000000028 | |
| Stark et al. (2015b) (MESSENGER laser altimeter and images) | 281.00980 ± 0.00088 | | 61.4156 ± 0.0016 | | | 6.13851804 ± 0.00000094 | 38.9 ± 1.3 |
| Stark (2016) (MESSENGER images) | | | | | **329.5988 ± 0.0037** | **6.13851079 ± 0.00000120** | **38.5 ± 1.6** |
| Verma and Margot (2016) (MESSENGER radio science) | 281.00975 ± 0.0048 | | 61.41828 ± 0.0028 | | | | |
| Baland et al. (2017) (based on Stark et al. (2015b)) | 281.00981 ± 0.00083 | −0.032907 | 61.41565 ± 0.00150 | −0.0048590 | | | |

Footnote:

Pettengill and Dyce (1965) measured a rotation period of 59±5 days based on radar echoes from Mercury's surface. Colombo (1965) computed a rotation period of 58.6462 days based on Mercury ephemeris. McGovern et al. (1965) obtained a rotation period of 58.4±0.4 days from examination of drawings of Mercury based on telescopic observations. Dyce et al. (1967) revised measurements by Pettengill and Dyce (1965) and obtained a rotation period of 59±3 days. Camichel and Dollfus (1968) and Smith and Reese (1968) used photographic measures to obtain a rotation period of 58.67 ± 0.03 days and 58.663 ± 0.021 days, respectively. The values denoted as IAU 1970 are computed from a transformation matrix reported by Davies and Batson (1975). Klaasen (1975) and (1976) analyzed Mariner 10 image data and reported a rotation period of 58.661 ± 0.017 days and 58.6461 ± 0.005 days, respectively. The values for the rotation axis orientation of Klaasen (1976) were obtained from reading the coordinates of Fig. 5 of his paper. Robinson et al. (1999) estimate an accuracy of about 20 km (0.47°) for their W0 estimate. Baland et al. (2017) have used the measurements of A. Stark, Oberst, Preusker, et al. (2015) at MJD56353.5 to provide an rotation axis orientation at the J2000.0 epoch, which is consistent with the assumption that Mercury occupies a Cassini state (the precession rates are obtained with the help of their obliquity estimate).

Table 2: Mercury shape characteristics and transformation parameters for the ellipsoid frame. A, B, C are the ellipsoid axes; (dx, dz, dy) denotes the Cartesian coordinates of the center of figure with respect to the center of mass; $\alpha, \beta, \gamma$ are Cardan angles for the transformation into the ellipsoid frame $E = R_x(\alpha) \cdot R_y(\beta) \cdot R_z(\gamma)$ .

|  | This work (MDIS DTM computed by Becker et al. (2016)) | MLA and MESSENGER radio link occultation (Perry et al., 2015) | Limb images (Elgner et al., 2014) | This work (average value of the first and second columns) |
| --- | --- | --- | --- | --- |
| A [km] | 2440.702 | 2440.53 ± 0.04 | 2441.737 ± 0.20 | 2440.616 ± 0.086 |
| B [km] | 2439.387 | 2439.28 ± 0.04 | 2439.537 ± 0.19 | 2439.334 ± 0.054 |
| C [km] | 2438.328 | 2438.26 ± 0.04 | 2439.377 ± 0.25 | 2438.294 ± 0.034 |
| R [km] | 2439.472 | 2439.36 ± 0.02 | 2440.287 ± 0.27 | 2439.416 ± 0.056 |
| dx [km] | 0.075 | 0.042 ± 0.030 | -0.0047 ± 0.160 | 0.059 ± 0.017 |
| dy [km] | 0.120 | 0.133 ± 0.040 | 0.1557 ± 0.159 | 0.127 ± 0.007 |
| dz [km] | -0.097 | -0.038 ± 0.040 | - | -0.0675 ± 0.030 |
| $\alpha$ [°] | 2.69 | -0.88 | 31.26 | 0.91 ± 1.79 |
| $\beta$ [°] | -3.42 | -1.92 | 1.70 | -2.67 ± 0.75 |
| $\gamma$ [°] | 16.52 | 15.07 | 26.66 | 15.8 ± 0.7 |